\documentclass[aps,prd,twocolumn]{revtex4}
\usepackage{graphicx}

\newcommand{\bel}[1]{\begin{equation}\label{#1}}
\newcommand{\bal}[1]{\begin{eqnarray}\label{#1}}

\newcommand{\be}{\begin{equation}}
\newcommand{\ee}{\end{equation}}
\newcommand{\ba}{\begin{eqnarray}}
\newcommand{\ea}{\end{eqnarray}}

\newcommand{\bes}{\begin{equation*}}
\newcommand{\ees}{\end{equation*}}

\begin{document}

\title{Chiral transition in a strong magnetic background}

\author{Eduardo S. {\sc Fraga}\footnote{fraga@if.ufrj.br} and 
Ana J\'ulia {\sc Mizher}\footnote{anajulia@if.ufrj.br} }

\affiliation{Instituto de F\'\i sica, Universidade Federal do Rio de Janeiro, \\
Caixa Postal 68528, Rio de Janeiro, RJ 21941-972, Brazil}

\begin{abstract}
The presence of a strong magnetic background can modify the nature and the 
dynamics of the chiral phase transition at finite temperature. We compute the 
modified effective potential in the linear sigma model with quarks to one loop 
in the $\overline{MS}$ scheme for $N_{f}=2$. For fields $eB\sim 5 m_{\pi}^{2}$ 
and larger a crossover is turned into a weak first-order transition. We discuss 
possible implications for non-central heavy ion collisions at RHIC and LHC, 
and for the primordial QCD transition.
\end{abstract}

\maketitle

\section{Introduction}

Strong magnetic fields can produce remarkable physical effects. One of the most 
spectacular examples is provided by magnetars \cite{magnetars}. 
More recently, it has been proposed that strong magnetic fields could play a very 
important role in the physics of high-energy heavy ion collisions, affecting observables 
in the case of non-central collisions at the Relativistic Heavy Ion Collider (RHIC), at 
Brookhaven, and the Large Hadron Collider (LHC), at CERN, and providing a possible 
signature of the presence of CP-odd domains in the presumably formed quark-gluon 
plasma (QGP) phase \cite{CP-odd}. In that case, one would reach magnetic fields 
$B\sim 10^{19}~$ Gauss, which correspond to $eB\sim 6 m_{\pi}^{2}$, where $e$ is 
the fundamental charge and $m_{\pi}$ is the pion mass. This gives a very intense magnetic 
background from the point of view of quantum chromodynamics (QCD) 
scales \footnote{For comparison, one has $m_{\pi}/e \sim 10^{19}~$Gauss.}.

In this paper we investigate the effects of a strong magnetic background on the nature and 
dynamics of the chiral phase transition at finite temperature, $T$, and vanishing chemical 
potential. As a framework, we adopt the linear sigma model coupled to quarks (LSM$_{q}$) 
with two flavors, $N_{f}=2$ \cite{GellMann:1960np}. This effective theory has been widely used 
to describe different aspects of the chiral transition, such as thermodynamic 
properties \cite{quarks-chiral,ove,Scavenius:1999zc,Caldas:2000ic,Scavenius:2000qd,
Scavenius:2001bb,paech,Mocsy:2004ab,Aguiar:2003pp,Schaefer:2006ds,Taketani:2006zg}
and the nonequilibrium phase conversion process \cite{Fraga:2004hp}, 
as well as combined to other models in order to include effects from 
confinement \cite{polyakov,explosive}.

In the limit of strong magnetic fields, we find that the nature of the chiral transition is 
modified. The original LSM$_{q}$, with no magnetic field, yields a crossover at 
$T=T_{c}\sim 150~$MeV for a choice of coupling to quarks that provides a reasonable 
mass for the constituent quarks at zero temperature \cite{Scavenius:2000qd}. 
The presence of a strong magnetic background turns this picture into a weak first-order 
phase transition. As a consequence, the dynamics that follows a rapid supercooling, 
such as the one that presumably happens in a high-energy heavy ion collision, can be 
dramatically modified due to the presence of a barrier in the effective potential. Even if 
the barrier is quite small, as in the case of a weak first-order transition, its effect can be 
very significant for the dynamics, holding the system in the false vacuum until it reaches 
the spinodal instability and explodes \cite{spinodal}.

From the theoretical point of view, the non-trivial role played by magnetic fields in the 
nature of phase transitions has been known for a long time \cite{landau-book}. Modifications 
in the vacuum of quantum electrodynamics (QED) and QCD have also been investigated 
within different frameworks, mainly using effective 
models \cite{Klevansky:1989vi,Gusynin:1994xp,Babansky:1997zh,Klimenko:1998su,
Semenoff:1999xv,Goyal:1999ye,Hiller:2008eh}, 
especially the NJL model \cite{Klevansky:1992qe}, and chiral perturbation 
theory \cite{Shushpanov:1997sf,Agasian:1999sx,Cohen:2007bt}, but also resorting to the 
quark model \cite{Kabat:2002er} and certain limits of QCD \cite{Miransky:2002rp}. 
Most treatments have been concerned with vacuum modifications by the magnetic field, 
though medium effects were considered in a few cases. More recently, interesting phases in 
dense systems \cite{Son:2007ny}, as well as effects on the dynamical quark 
mass \cite{Klimenko:2008mg} and on the quark-hadron transition \cite{Agasian:2008tb} 
were also considered. 

The paper is organized as follows. Section II presents briefly the 
low-energy effective model adopted in this paper, as well as a discussion 
of the approximations required. In Section III we show our results for 
the modified effective potential, in the presence of a strong magnetic 
background. Phenomenological consequences are illustrated 
and discussed in Section IV. Section V contains our conclusions and outlook. 

\section{Effective theory}

To describe the chiral phase structure of strong interactions at finite 
temperature we adopt the LSM$_{q}$, defined by the following lagrangian
\begin{eqnarray}
{\cal L} &=&
 \overline{\psi}_f \left[i\gamma ^{\mu}\partial _{\mu} - g(\sigma +i\gamma _{5}
 \vec{\tau} \cdot \vec{\pi} )\right]\psi_f \nonumber\\
&+& \frac{1}{2}(\partial _{\mu}\sigma \partial ^{\mu}\sigma + \partial _{\mu}
\vec{\pi} \partial ^{\mu}\vec{\pi} )
- V(\sigma ,\vec{\pi})\;,
\label{lagrangian}
\end{eqnarray}
where
\begin{equation} 
V(\sigma ,\vec{\pi})=\frac{\lambda}{4}(\sigma^{2}+\vec{\pi}^{2} -
{\it v}^2)^2-h\sigma
\label{bare_potential}
\end{equation}
is the self-interaction potential for the mesons, exhibiting both spontaneous 
and explicit breaking of chiral symmetry. The $N_f=2$ massive fermion fields 
$\psi_f$ represent the up and down constituent-quark fields $\psi=(u,d)$. The 
scalar field $\sigma$ plays the role of an approximate order parameter for the 
chiral transition, being an exact order parameter for massless quarks and pions. 
The latter are represented by the pseudoscalar field $\vec{\pi}=(\pi^{0},\pi^{+},\pi^{-})$, 
and it is common to group together these meson fields into an $O(4)$ chiral field 
$\phi =(\sigma,\vec{\pi})$. 

In what follows, we implement a simple mean-field treatment with the customary simplifying 
assumptions (see, e.g., Ref. \cite{Scavenius:2000qd}). Quarks constitute a thermalized fluid 
that provides a background in which the long wavelength modes of the chiral condensate evolve. 
Hence, at $T=0$, the model reproduces results from the usual LSM without 
quarks or from chiral perturbation theory for the broken phase vacuum \cite{FTFT-books}. 
In this phase, quark degrees of freedom are absent (excited only for $T > 0$). The $\sigma$ 
field is heavy, $M_{\sigma} \sim600$ MeV, and treated classically. On the other hand, pions are 
light, and fluctuations in $\pi^{+}$  and $\pi^{-}$ couple to the magnetic field, $B$, as will be 
discussed in the next section, whereas fluctuations in $\pi^{0}$ give a $B$-independent 
contribution that we ignore, for simplicity. For $T > 0$, quarks are relevant (fast) degrees of 
freedom and chiral symmetry is approximately restored in the plasma for high enough $T$. 
In this case, we incorporate quark thermal fluctuations in the effective potential for $\sigma$, 
i.e. we integrate over quarks to one loop. Pions become rapidly heavy only after $T_{c}$ and 
their fluctuations can, in principle, matter since they couple to $B$. However, they will play a 
minor role in this case, as will be clear later, so that we ignore their thermal fluctuations in 
this section, for simplicity.

The parameters of the lagrangian are chosen such that the effective model reproduces 
correctly the phenomenology of QCD at low energies and in the vacuum, in the absence of 
a magnetic field, such as the spontaneous (and small explicit) breaking of chiral 
symmetry and experimentally measured meson masses. So, one has to impose that 
the chiral $SU_{L}(2) \otimes SU_{R}(2)$ symmetry is spontaneously broken in the 
vacuum, and the expectation values of the condensates are given by 
$\langle\sigma\rangle ={\it f}_{\pi}$ and $\langle\vec{\pi}\rangle =0$, where 
${\it f}_{\pi}=93$~MeV is the pion decay constant. The explicit symmetry breaking 
term is determined by the PCAC relation which gives $h=f_{\pi}m_{\pi}^2$, where
$m_{\pi}\approx 138~$MeV is the pion mass. This
yields $v^{2}=f^{2}_{\pi}-{m^{2}_{\pi}}/{\lambda}$. The value of 
$\lambda = 20$ leads to a $\sigma$-mass, 
$m^2_\sigma=2\lambda f^{2}_{\pi}+m^{2}_{\pi}$, equal to $600~$MeV. 
In mean field theory, the purely bosonic part of this Lagrangian 
exhibits a second-order phase transition~\cite{Pisarski:1984ms} 
at $T_c=\sqrt{2}v$ if the explicit symmetry breaking term, $h_q$, 
is dropped. For $h_q\ne 0$, the transition becomes a smooth crossover 
from the restored to broken symmetry phases. For $g>0$, one has 
to include a finite-temperature one-loop contribution from the quark 
fermionic determinant to the effective potential. The choice $g=3.3$, which 
yields a reasonable mass for the constituent quarks in the vacuum, 
$M_{q}= g f_{\pi}$, leads 
to a crossover at finite temperature and vanishing chemical potential, and will 
be adopted in this paper \footnote{It is also common to use $g=5.5$ when one 
aims to describe a first-order chiral transition. However, since the constituent quark 
mass is given by $M_{q}=g\langle\sigma\rangle$, this choice yields a constituent 
quark mass that is too high.}.

Standard integration over the fermionic degrees of freedom \cite{FTFT-books}, 
using a classical approximation for the chiral field, gives the following formal expression 
for the effective potential in the $\sigma$ direction:
\begin{equation}
V_{eff}(T,\sigma)= V(\sigma) - \frac{T}{{\cal V}}
\ln\det\left[\frac{(G_E^{-1}+M_{q}(\sigma))}{T}\right]\; ,
\label{TDpot}
\end{equation}
where $V(\sigma)$ is the classical self-interaction potential for the mesonic sector in 
the $\sigma$ direction, $G_E$ is the fermionic Euclidean propagator, $M_{q}$ 
is the effective fermion mass in the presence of the chiral field background, 
and ${\cal V}$ is the volume of the system. To one loop, one has $V_{eff}= V(\phi)+V_q(\phi)$, 
where the contribution from the quarks reads
\begin{equation}
V_q\equiv -\nu_q T \int \frac{d^3k}{(2\pi)^3} \ln\left( 1 + e^{-E_k(\sigma)/T} \right) \; .
\end{equation}
where $\nu_q=24$ is the color-spin-isospin-baryon charge degeneracy factor, 
$E_k(\sigma)=(\vec{k}^2+M_{q}^2(\sigma))^{1/2}$, and 
$M(\sigma)=g|\sigma|$ plays the role of an effective mass for the quarks. 
The net effect of this term is correcting the potential for the chiral 
field, approximately restoring chiral symmetry for a critical temperature 
$T_{c}\sim 150~$MeV \cite{Scavenius:2000qd}. The effective potential $V_{eff}$ 
for several values of the temperature is displayed in Fig. 1, assuming $g=3.3$. 
\vspace{0.6cm}
\begin{figure}[htb]
\includegraphics[width=7.7cm]{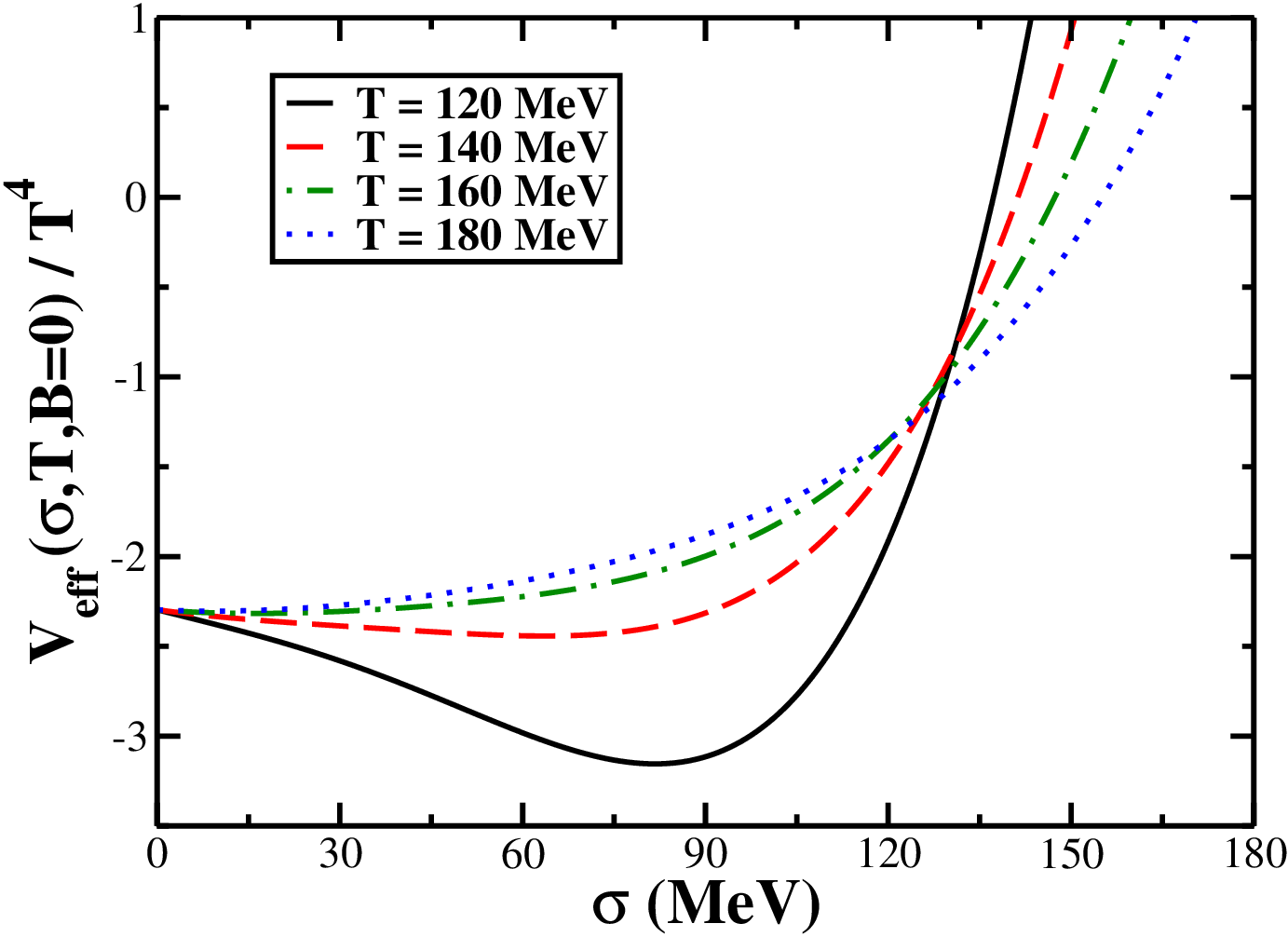}
\caption{$V_{eff}(\sigma)$ for different values of the temperature 
for $g=3.3$, in the absence of a magnetic field.}
\label{Veff}
\end{figure}
%

\section{Modified effective potential}

Let us now assume that the system is in the presence of a strong magnetic field background 
that is constant and homogeneous. For definiteness, let us take the direction of the magnetic 
field as the $z$-direction, $\vec{B}=B \hat z$. One can compute the modified effective potential 
by redefining the dispersion relations of the scalar and spinor fields in the presence of $\vec{B}$, 
using the minimal coupling shift in the gradient and the field equations of motion. For this purpose, 
it is convenient to choose the gauge such that $A^{\mu}=(A^{0},\vec{A})=(0,-By,0,0)$.

For scalar fields with electric charge $q$, one has
\begin{eqnarray}
&&(\partial^2 +m^2)\phi=0 \, ,\\
&&\partial_\mu \to \partial_\mu+iq A_\mu \, .
\end{eqnarray}
After decomposing $\phi$ into Fourier modes, except for the dependence in the 
coordinate $y$, one obtains
\begin{eqnarray}
\varphi''(y)&+&2m\left[ \left( \frac{p_0^2-p_z^2-m^2}{2m} \right) \right. \nonumber \\
&-& \left. \frac{q^2B^2}{2m}\left( y+\frac{p_x}{qB} \right)^2\right]\varphi(y)=0 \, ,
\end{eqnarray}
which has the form of a Schr\"odinger equation for a harmonic oscillator. Its eigenmodes 
correspond to the well-known Landau levels
\begin{eqnarray}
\varepsilon_n &\equiv& \left( \frac{p_{0n}^2-p_z^2-m^2}{2m} \right)=
\left( n+\frac{1}{2} \right)\omega_B \, ,
\end{eqnarray}
where $\omega_B = |q|B/m$ and $n$ is an integer, and provide the new dispersion relation:
\begin{eqnarray}
p_{0n}^2=p_z^2+m^2+(2n+1)|q|B \, .
\end{eqnarray}

One can proceed in an analogous way for fermions with charge $q$. From the free Dirac 
equation $(i \gamma^\mu\partial_\mu -m)\psi=0$, and the shift in $\partial_{\mu}$, 
one arrives at the following Schr\"odinger equation
\begin{eqnarray}
u_{\sigma}''(y)&+&2m \left[ \left( \frac{p_0^2-p_z^2-m^2+qB\sigma}{2m} \right) \right. \nonumber \\
&-& \left. \frac{q^2B^2}{2m}\left( y+\frac{p_x}{qB} \right)^2\right]u_{\sigma}(y)=0 \, ,
\end{eqnarray}
which yields the new dispersion relation for quarks:
\begin{eqnarray}
p_{0n}^2=p_z^2+m^2+(2n+1-\sigma)|q|B \, .
\end{eqnarray}

It is also straightforward to show that integrals over four momenta and thermal 
sum-integrals acquire the following forms, respectively:
\begin{eqnarray}
&&\int \frac{d^4k}{(2\pi)^4} \mapsto \frac{|q|B}{2\pi}\sum_{n=0}^\infty 
\int \frac{dk_0}{2\pi}\frac{dk_z}{2\pi} \, ,\\
T\sum_{\ell} &&\int \frac{d^3k}{(2\pi)^3} \mapsto \frac{|q|BT}{2\pi}\sum_\ell \sum_{n=0}^\infty 
\int \frac{dk_z}{2\pi} \, ,
\end{eqnarray}
where $n$ represents the different Landau levels and $\ell$ stands for the Matsubara 
frequency indices \cite{FTFT-books}.

According to the assumptions of our effective model, the vacuum potential will be 
modified by a contribution from the charged pions that couple to the magnetic field. To one 
loop, this is given by the following integral:
\begin{eqnarray}
V_{\pi^\pm}^V=\frac{1}{2}\frac{|q|B}{2\pi}\sum_{n=0}^\infty 
\int_{-\infty}^{+\infty} \frac{dk}{2\pi}\left[ k^2+m_\pi^2+(2n+1)|q|B \right]^{1/2}
\, .\nonumber \\
\end{eqnarray}
For sufficiently high magnetic fields, $eB >> m_{\pi}^{2}$, and ignoring contributions 
independent of the chiral condensate, this simplifies to the following renormalized 
result \footnote{High magnetic fields allow for an expansion of generalized zeta functions, 
of the form $\zeta\left( -1 +\frac{\epsilon}{2},\frac{m_\pi^2+|q|B}{2|q|B} \right)$, that appear 
in the $\overline{MS}$ computation of the momentum integral.}:
\begin{eqnarray}
V_{\pi^+}^V+V_{\pi^+}^V=-\frac{2m_\pi^2 eB}{32\pi^2}\log 2 \, ,
\end{eqnarray}
where we have added the contributions from $\pi^{+}$ and $\pi^{-}$.

\vspace{0.6cm}
\begin{figure}[htb]
\includegraphics[width=7.7cm]{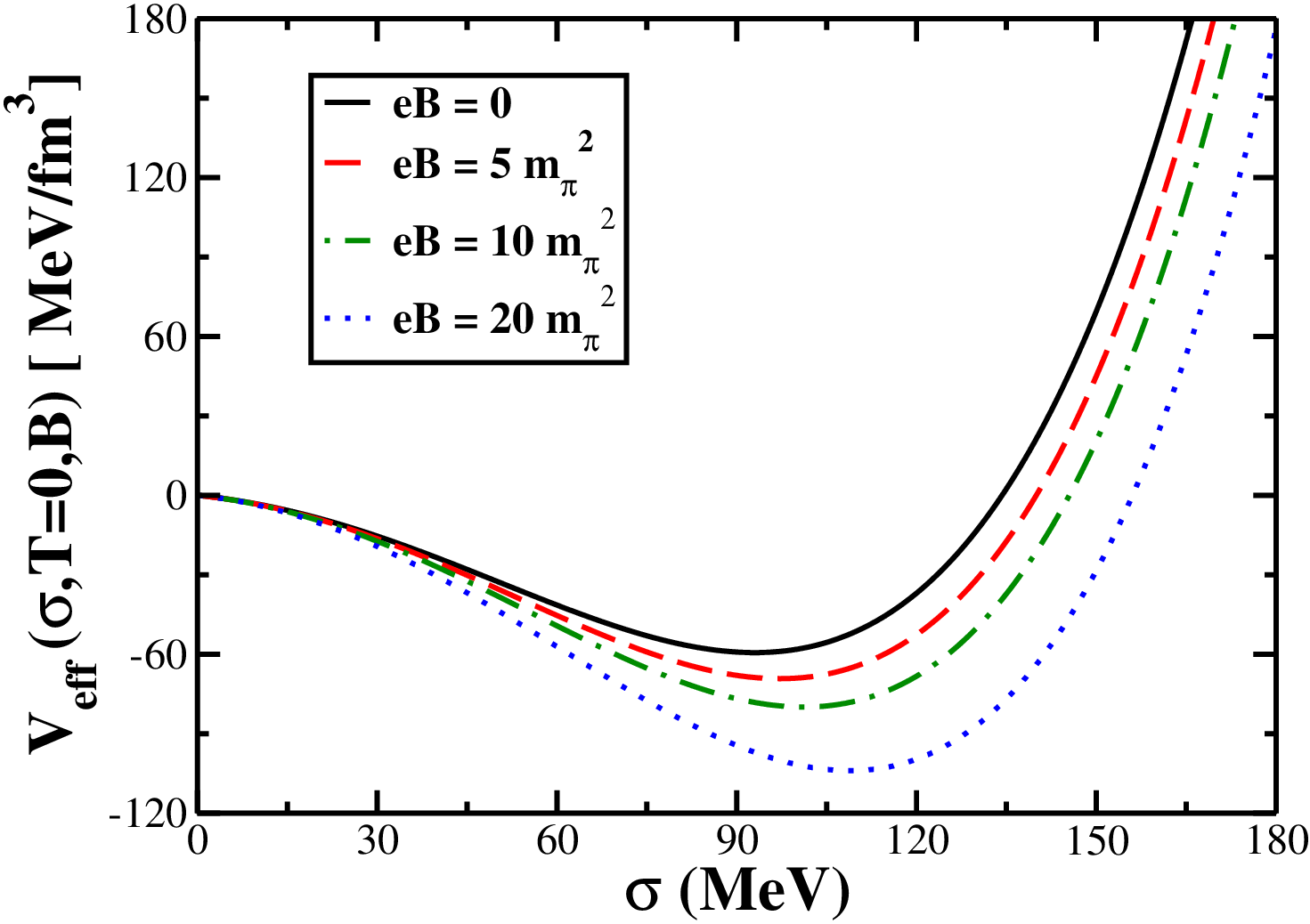}
\caption{$V_{eff}(\sigma,T,B)$ for $T=0$ and different values of the magnetic field.}
\label{Veff_Bvac}
\end{figure}

Results for the modified vacuum potential are displayed in Fig. 2 for different values 
of the background magnetic field. As one can see from the plot, the presence of the 
magnetic field enhances the value of the chiral condensate and the depth of the 
broken phase minimum of the modified effective potential, a result that is in line 
with those found within different approaches (see, for instance, 
Refs. \cite{Shushpanov:1997sf,Cohen:2007bt,Hiller:2008eh}).
This could be an indication that the presence of a strong magnetic background 
might favor a weak first-order transition over the crossover that is obtained for 
$B=0$. To investigate this point, one has to compute the modified thermal contributions 
to the effective potential and study the process of chiral symmetry restoration.

Charged pions contribute to the one-loop thermal corrections via
\begin{eqnarray}
V_{\pi^\pm}^T=\frac{|q|BT}{2\pi}\sum_{n=0}^\infty \int_{-\infty}^{+\infty} 
\frac{dk}{2\pi}\log\left( 1 -e^{-\beta\omega_{\pi n}} \right) \, ,
\end{eqnarray}
whereas the contribution from quarks reads

\begin{eqnarray}
V_{q}^T=-\sum_{\sigma=\pm}\frac{|q|BT}{2\pi}\sum_{n=0}^\infty 
\int_{-\infty}^{+\infty} \frac{dk}{2\pi}2\log\left( 1 + e^{-\beta\omega_{n\sigma}} \right) \, ,
\end{eqnarray}
where, for quarks, one still has to sum over colors, which brings a factor $N_{c}$, and over 
flavors. The latter sum, differently from what happened in the previous section, does 
not bring and overall $N_{f}$ factor, since the electric charges are not equal, and 
their absolute values add up to one. In the presence of a nonzero $B$, spins are also 
treated separately, as is clear in the expression above.

\vspace{0.6cm}
\begin{figure}[htb]
\includegraphics[width=7.7cm]{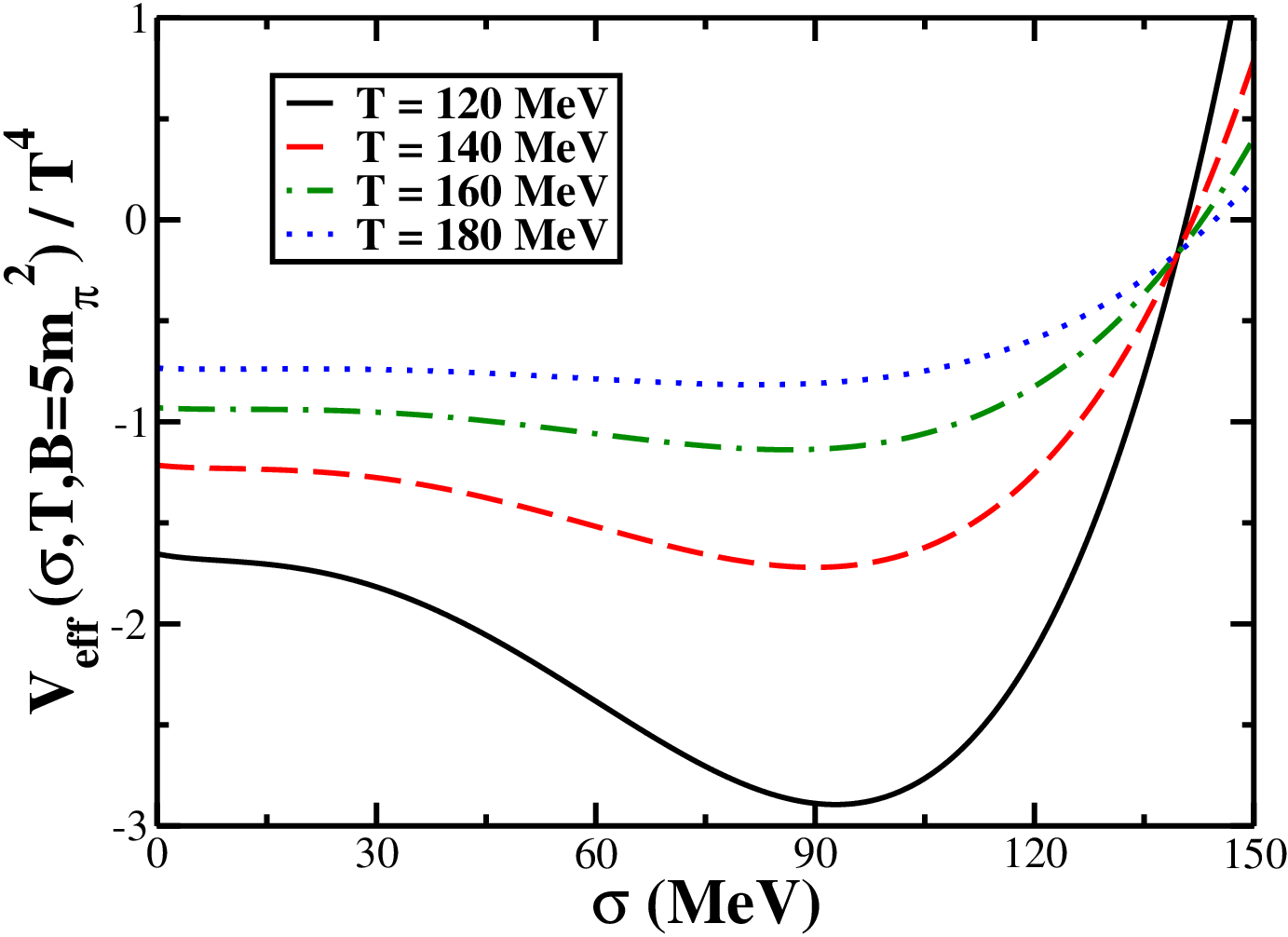}
\caption{$V_{eff}(\sigma,T,B)$ for $eB=5 m_{\pi}^{2}$ and different values of the temperature.}
\label{Veff_B1}
\end{figure}

For sufficiently high magnetic fields, not only compared to the pion mass but also to the 
effective quark mass and to the temperature, and again ignoring contributions independent 
of the chiral condensate, we obtain for charged pions and quarks, respectively:
\begin{eqnarray}
V_{\pi^+}^T+V_{\pi^-}^T=-2\frac{(eB)^2}{(2\pi)^{3/2}}
\left( \frac{T^2}{eB} \right)^{3/4} e^{-\beta\sqrt{eB}} 
\left[ 1 + \frac{m_\pi^2}{4eB} \right] \, , \nonumber \\
\end{eqnarray}
\begin{eqnarray}
V_q^T&=&-N_c\frac{eBT^2}{2\pi^2}\left[ \int_{-\infty}^{+\infty} dx 
\log \left( 1 + e^{-\sqrt{x^2+m_f^2/T^2}} \right) \right. \nonumber \\
&+& \left. \sqrt{2\pi\beta\sqrt{2eB}}~e^{-\beta\sqrt{2eB}} 
\left( 1+ \frac{m_f^2}{8eB} \right) \right] \, .
\end{eqnarray}
\vspace{0.6cm}
\begin{figure}[htb]
\includegraphics[width=7.7cm]{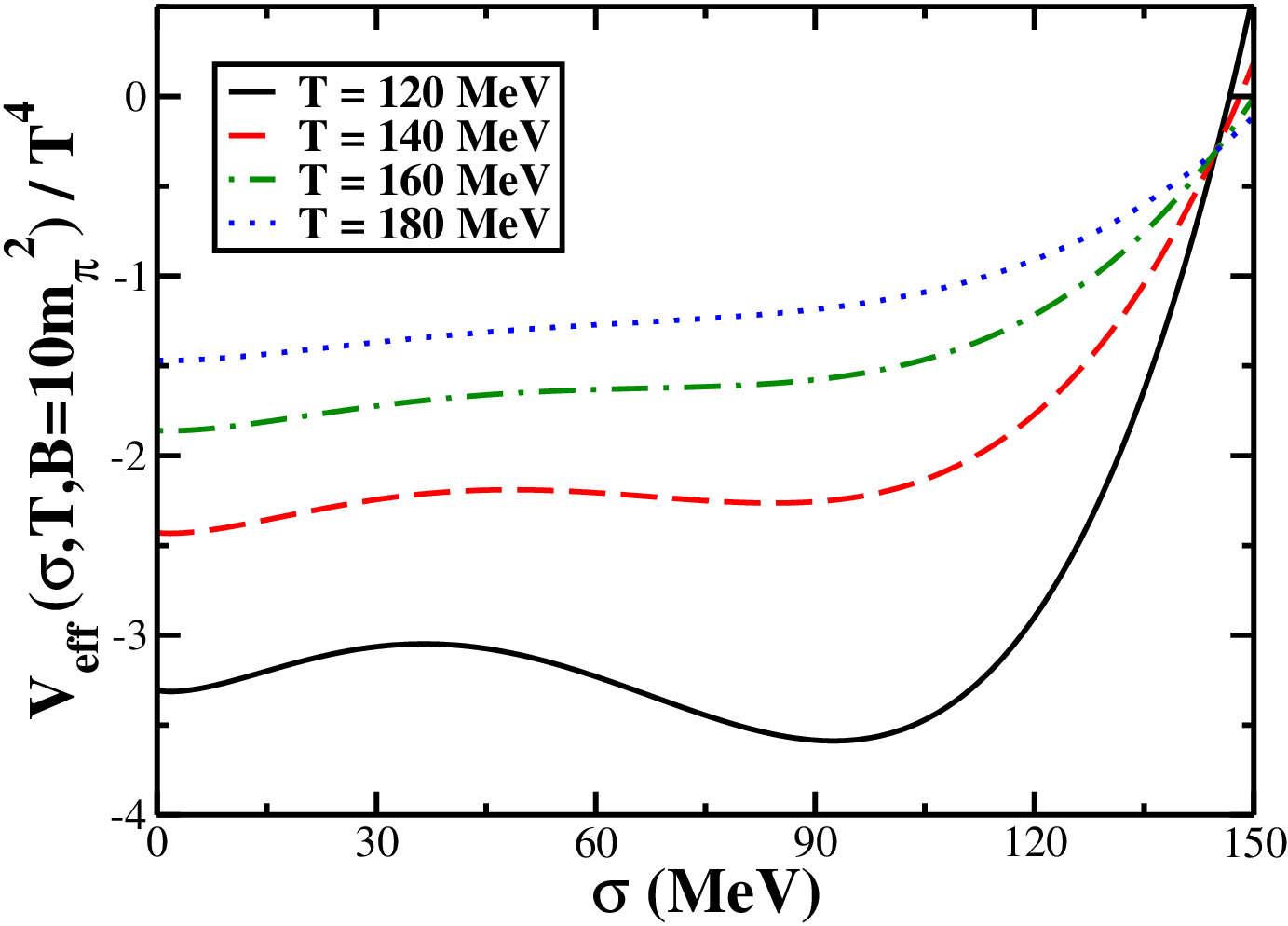}
\caption{$V_{eff}(\sigma,T,B)$ for $eB=10 m_{\pi}^{2}$ and different values of the temperature.}
\label{Veff_B2}
\end{figure}

One can notice that the pion thermal contribution and part of the quark thermal 
contribution are exponentially suppressed for high magnetic fields, which 
essentially comes from an increase in the effective mass due to the magnetic 
field $\sim e^{-m_{eff}/T}\sim e^{-\#\sqrt{eB}/T}$. Therefore, within these 
approximations, the total modified effective potential is given by
\begin{eqnarray}
&&V_{eff}(\sigma,T,B)\approx \frac{\lambda}{4}(\sigma^2-v^2)^2-h\sigma  - 
2\frac{m_\pi^2eB}{32\pi^2}\log 2 \nonumber \\
&-& 2N_c\frac{eBT^2}{2\pi^2}\int_{0}^{\infty} dx 
\log \left( 1 + e^{-\sqrt{x^2+m_f^2/T^2}} \right) \, . 
\end{eqnarray}

The modified effective potential, $V_{eff}(\sigma,T,B)$, is plotted for different temperatures 
for $eB=5 m_{\pi}^{2}$, $eB=10 m_{\pi}^{2}$ and $eB=20 m_{\pi}^{2}$ in Fig. 3, Fig. 4 and 
Fig. 5, respectively. For $eB=5 m_{\pi}^{2}$, one verifies that the critical temperature becomes 
higher, $T_{c}> 200~$MeV. Moreover, a closer look uncovers the presence of a tiny barrier, 
signaling a first-order phase transition.  A zoom for a couple of temperatures close to $T_{c}$ 
is shown in Fig. 6. In the case of higher magnetic fields, the existence of the barrier is rather 
clear, showing that the presence of a strong magnetic background can really modify the nature 
of the chiral transition, in this case from a crossover to a first-order transition whose strength 
depends on how intense $B$ is. In Figs. 4, 5 and 6, we considered the same set of temperatures 
to allow for a more direct comparison.

\vspace{0.6cm}
\begin{figure}[htb]
\includegraphics[width=7.7cm]{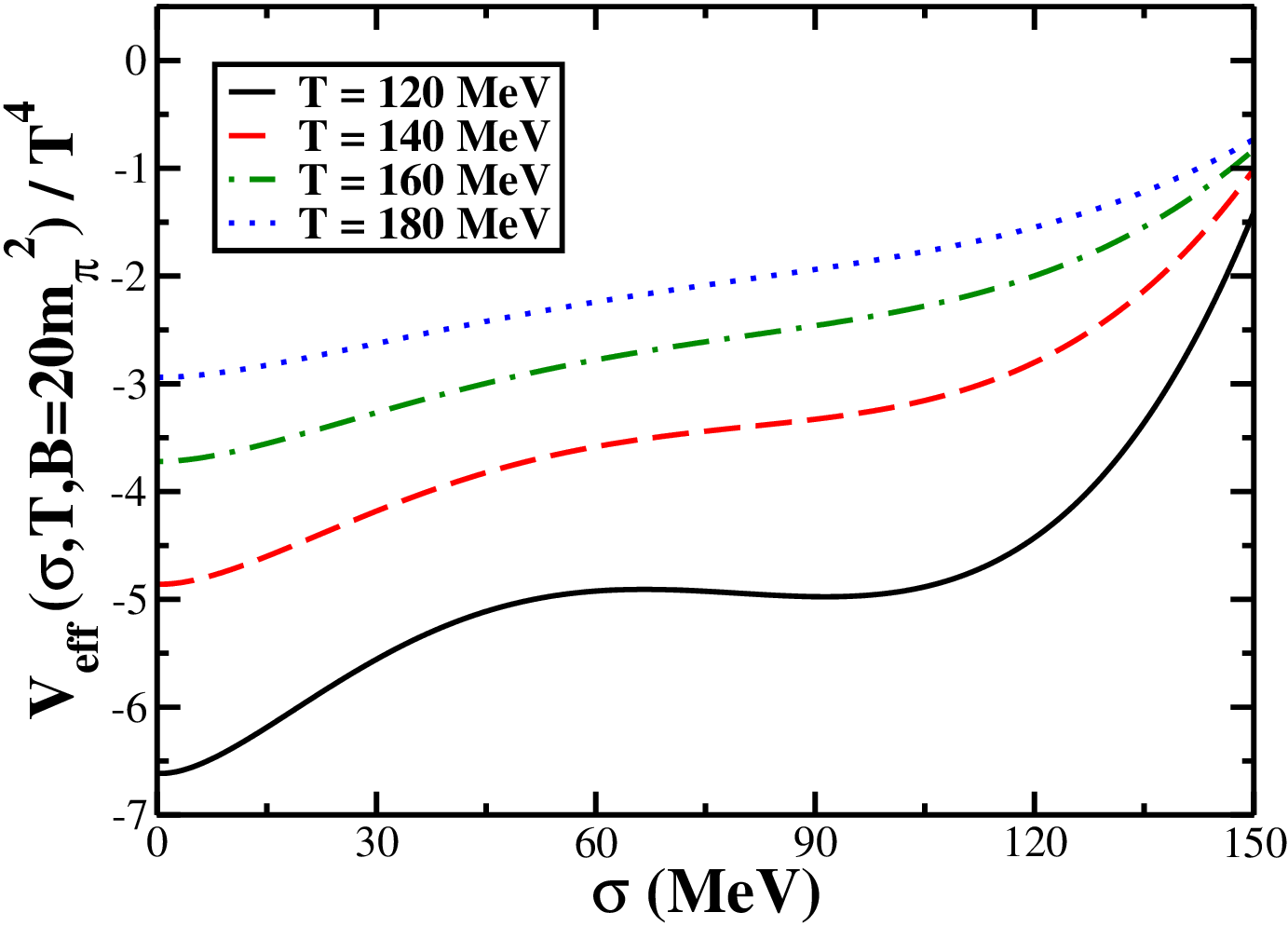}
\caption{$V_{eff}(\sigma,T,B)$ for $eB=20 m_{\pi}^{2}$ and different values of the temperature.}
\label{Veff_B3}
\end{figure}

Another important feature that comes out of this comparison is the fact that for $eB=10 m_{\pi}^{2}$ 
the critical temperature drops again ($T_{c}< 140~$MeV) as compared to the case in 
which $eB=5 m_{\pi}^{2}$. This phenomenon continues to happen for higher $B$ (see 
Fig. 6), showing that there is a non-trivial balance between temperature and magnetic field 
that can bring some richness to the chiral $T-B$ phase diagram that should be explored. 
The presence of the barrier is not qualitatively modified, though, and the ``conversion'' to a 
first-order transition seems to be robust.

\vspace{0.6cm}
\begin{figure}[htb]
\includegraphics[width=7.7cm]{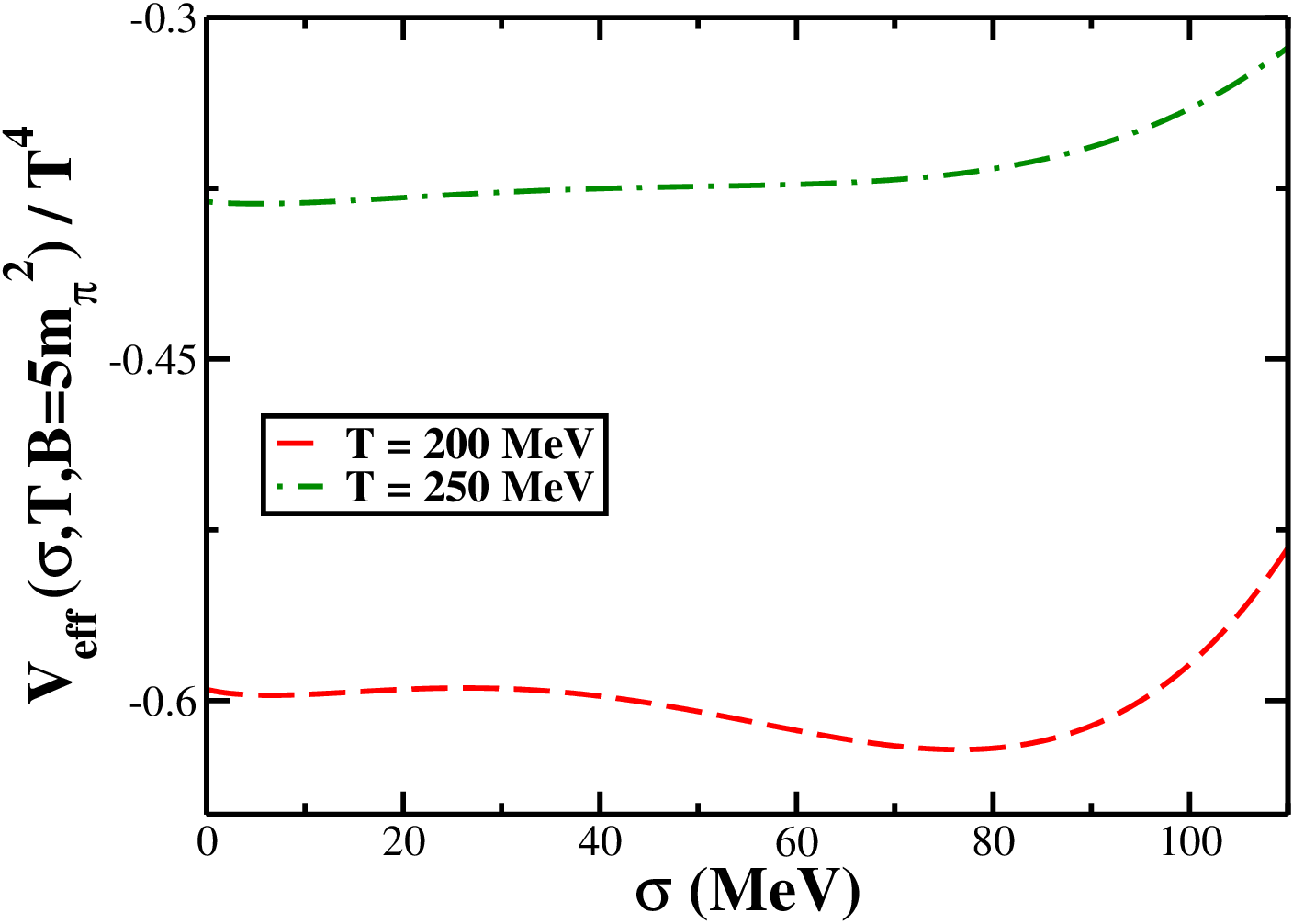}
\caption{$V_{eff}(\sigma,T,B)$ for $eB=5 m_{\pi}^{2}$ and different values of the temperature 
(zoom near the critical temperature to exhibit the barrier).}
\label{Veff_B1_zoom}
\end{figure}
%

\section{Phenomenological consequences}

Besides the pure theoretical interest in the modification of the nature of the chiral phase 
transition by the presence of a strong magnetic background, one should consider possible 
phenomenological implications within the reach of present experiments. At RHIC, estimates 
by Kharzeev, McLerran and Warringa \cite{CP-odd} give $eB\sim 5.3 m_{\pi}^{2}$. For the 
LHC, there is a factor $Z_{Pb}/Z_{Au} = 82/79$ coming from the difference in charge, 
and some small increase in the maximum 
value of $eB$ due to the higher center-of-mass energy (as observed for RHIC \cite{CP-odd}). 
So, it is reasonable to consider $eB\sim 6 m_{\pi}^{2}$ as a representative estimate.

We show in Fig. 7 a zoom of the effective potential for $eB\sim 6 m_{\pi}^{2}$ for a temperature 
slightly below the critical one, whereas the general picture is not appreciably different from 
the case $eB\sim 5 m_{\pi}^{2}$, and can be seen in Fig. 3. Comparing Figs. 1 and 3, one 
can establish the basic differences brought about a magnetic field of the magnitude that could 
possibly be found in non-central high-energy heavy ion collisions. One moves from a crossover 
scenario to that of a weak first-order chiral transition, with a critical temperature 
$\sim 30\%$ higher.

In the case of a crossover, one expects the system to be smoothly drained to the true vacuum, 
with no formation of bubbles or spinodal decomposition. In the case of the first order transition 
depicted in Fig. 3, despite the fact that the barrier is quite small, part of the system will be kept in 
the false vacuum for a while, a few bubbles will be formed, depending on the intensity of the 
supercooling, and spinodal instability will then set in \cite{spinodal}.

The case of the LSM$_{q}$ with a coupling constant $g = 5.5$ exhibits a first-order phase 
transition even for $B=0$, and has been often used to study nucleation and spinodal 
decomposition phenomena in the chiral 
transition \cite{Scavenius:1999zc,explosive,Scavenius:2001bb,paech,Aguiar:2003pp,
Taketani:2006zg,Fraga:2004hp}. 
There, one finds a barrier $\sim 0.25$ 
in the effective potential for temperatures close to $T_{c}$, in units of $T^{4}$, and it has 
been shown that the system stays mostly apprisionated in the false vacuum until it reaches 
the spinodal explosive phase conversion regime \cite{Scavenius:2001bb}. 
For $g = 3.3$, we saw that there is a very weak 
first-order chiral transition for $ B > 0$, with a barrier an order of magnitude smaller (see 
Fig. 7). Nevertheless, even such small barriers can hold the system in the false vacuum 
until the spinodal instability in the case of a fast supercooling, as expected for high-energy 
heavy ion collisions \cite{spinodal}. Therefore, the existence of a strong magnetic background 
can bring significant effects to the dynamics of the chiral transition.

\vspace{0.6cm}
\begin{figure}[htb]
\includegraphics[width=7.7cm]{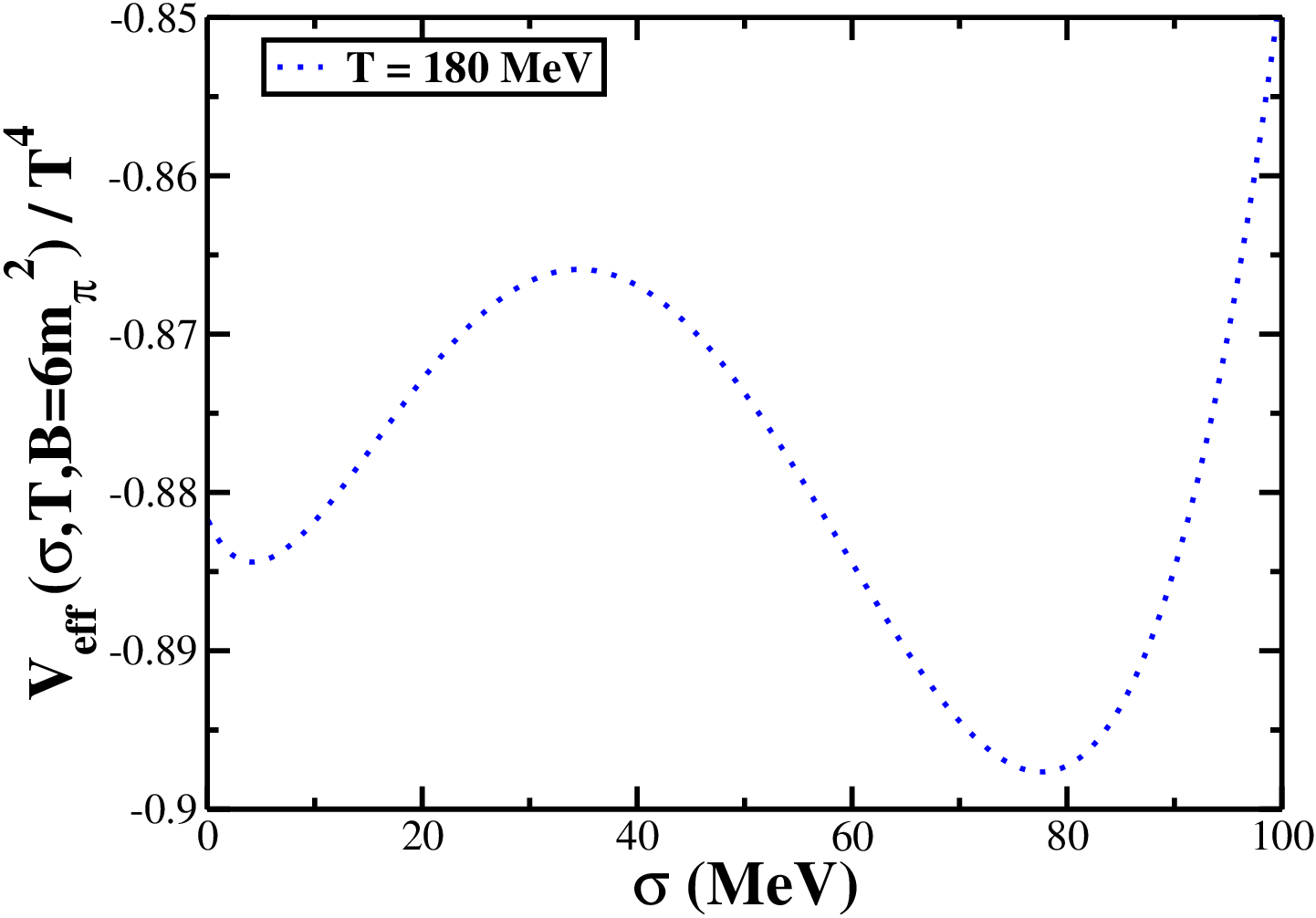}
\caption{$V_{eff}(\sigma,T,B)$ for $eB=6 m_{\pi}^{2}$ for $T=180~$MeV.}
\label{Veff_B_6_zoom}
\end{figure}
%

\section{Conclusions and outlook}

Lattice QCD indicates a crossover instead of a first-order chiral transition at finite temperature 
and vanishing chemical potential. However, a strong magnetic background might modify this 
situation. Our computation of the modified effective potential clearly shows the formation of a barrier 
that seems to be robust, in the case of magnetic fields that are larger than the other relevant 
scales, namely the pion mass and the temperature. On the other hand, it is clear that further 
investigation of the low magnetic field regime at finite temperature, for both $B < T$ and $B \sim T$, 
in the context of the chiral transition is called for, since the phase diagram seems to exhibit 
non-trivial features in this region of parameters which is interesting for critical phenomena 
in strong interactions.

For heavy ion collisions at RHIC and LHC, the barrier in the effective potential seems to be 
quite small. Nevertheless, it can probably hold most of the system in a metastable state down 
to the spinodal explosion, implying a different dynamics of phase conversion as compared 
to the crossover scenario. Moreover, the phenomenology resulting from varying $T$ and $B$ 
seems to be rich: there seems to be a competition between strengthening the chiral symmetry 
breaking via vacuum effects and its restoration by the thermal (magnetic) bath. In particular, 
non-central heavy ion collisions might show features of a first-order transition when contrasted 
to central collisions.

The dynamics of the chiral phase transition will certainly be affected by the presence of a 
strong magnetic background, since the nature of the transition is modified. In order to 
study its impact on the relevant time scales of this process, one has to perform real-time 
simulations of the evolution of the order parameter in the presence of the modified effective 
potential, e.g., in a Langevin framework. Results in this direction will be presented 
elsewhere \cite{future}.

Of course, the description presented in this paper is admittedly very simple. In actual heavy 
ion collisions, the magnetic field is neither uniform nor constant in time, and the values quoted 
here represent estimates of its maximum magnitude. In fact, the field intensity rapidly decreases 
with time, and any of the effects discussed here should manifest in the early-time dynamics. 
Moreover, if there is a a fast variation of $B$ with time, there will be an induced electric field 
that could break the vacuum through the Schwinger mechanism of pair production, and this 
would modify the condensate \cite{Cohen:2007bt} and the vacuum effective potential. 
Therefore a more realistic description of the external electromagnetic field effects on the chiral 
transition must be pursued before one is able to make explicit phenomenological predictions. 
On the other hand, we believe that the results presented here, even if seen as just qualitative, 
are encouraging. In particular, the possibility that strong magnetic fields can turn a crossover 
into a first-order transition might be relevant for the physics of the primordial QCD transition, 
eventually reconciling lattice results that predict a crossover \cite{Laermann:2003cv} 
to the description of phase conversion via bubble nucleation that assume a first-order 
phase transition \cite{Schwarz:2003du}.

\section*{Acknowledgments} 
We thank J.-B. Blaizot and A. Starinets for discussions. 
The authors would also like to thank E. Iancu for his kind hospitality at the 
Service de Physique Th\'eorique, CEA/Saclay, where this work has been concluded. 
This work was partially supported by CAPES-COFECUB, CNPq, FAPERJ and FUJB/UFRJ.


\end{document}